\documentclass[prb,twocolumn,amsmath,showpacs]{revtex4}
\usepackage{graphicx}

\begin{document}

\bibliographystyle{prsty}

\title {Bulk antiferromagnetism in $\bf Na_{0.82}CoO_2$ single crystals}

\author {S. Bayrakci$^1$,
 C. Bernhard$^1$,
 D. P. Chen$^1$,
 B. Keimer$^1$,
 R. K. Kremer$^1$,
 P. Lemmens$^1$,
 C. T. Lin$^1$,
 C. Niedermayer$^2$,
 and J. Strempfer$^1$}

\affiliation{$^1$Max Planck Institute for Solid State Research,
Heisenbergstr. 1, D-70569 Stuttgart, Germany\\
$^2$Paul-Scherrer-Institut, CH-5232 Villigen, Switzerland}\

\date{\today}

\begin{abstract}
Susceptibility, specific heat, and muon spin rotation measurements
on high-quality single crystals of $\rm Na_{0.82}CoO_2$ have
revealed bulk antiferromagnetism with N\'{e}el temperature $\rm
T_N = 19.8 \pm 0.1$ K and an ordered moment perpendicular to the
$\rm CoO_2$ layers. The magnetic order encompasses nearly 100\% of
the crystal volume. The susceptibility exhibits a broad peak
around 30 K, characteristic of two-dimensional antiferromagnetic
fluctuations. The in-plane resistivity is metallic at high
temperatures and exhibits a minimum at $\rm T_N$.
\end{abstract}

\pacs{75.30.-m, 76.75.+i, 72.80.Ga, 71.30.+h} \maketitle

Long-standing interest in the interplay between spin, charge, and
orbital degrees of freedom in cobalt oxides has increased
following the recent discovery of superconductivity in the
hydrated cobaltate $\rm Na_{0.35}CoO_2 \cdot 1.3\, H_2O$
(Ref.~\onlinecite{takada03}). Its transition temperature, $\rm
T_c=4.6~K$, is much lower than those of the layered copper oxides,
but higher than that of $\rm Sr_2 RuO_4$, the only other known
layered transition metal oxide which exhibits superconductivity.
The crystal structure of this compound contains distorted,
edge-sharing $\rm CoO_6$ octahedra such that the Co sites form a
triangular lattice \cite{lynn03,jorgensen03}. The electronic
structure is quasi-two-dimensional
\cite{singh00,koshibae03,honerkamp03,renner03}. Unconventional
superconductivity has been proposed based on NMR experiments
\cite{waki03,fujimoto03} and model calculations
\cite{singh00,baskaran03,ogata03,motrunich03}, but many important
questions about the symmetry of the order parameter and the
character of the quasi-particles at the Fermi surface have yet to
be answered.

The triangular $\rm CoO_2$ layers of $\rm Na_{x}CoO_2$ are
isostructural to those of $\rm Na_{0.35}CoO_2 \cdot 1.3\, H_2O$
(Refs. \onlinecite{fouassier73,jansen74,balsys96}). It may
therefore serve as a reference system with respect to the effects
of doping and magnetic correlations. $\rm Na_{x}CoO_2$ is metallic
\cite{tanaka94} and exhibits an unusual Hall effect \cite{wang03},
as well as a very large thermopower \cite{terasaki97}. The
thermopower increases with increasing x \cite{motohashi01}, which
in a local-moment picture corresponds to a progressive dilution of
magnetic $\rm Co^{4+}$ ($S=1/2$) with nonmagnetic $\rm Co^{3+}$
($S=0$). Following observations of a pronounced field dependence,
a magnetic mechanism has been invoked to explain the large
thermopower \cite{wang03}. It is therefore important to determine
the magnetic ground state and exchange parameters in this
material.

We have taken a first step in this direction by synthesizing
high-quality single crystals of $\rm Na_{x}CoO_2$ with x$\sim 0.8$
and studying their resistivity, specific heat, and magnetic
susceptibility as a function of temperature. An antiferromagnetic
state with N\'{e}el temperature $\rm T_N = 19.8 \pm 0.1$ K and an
ordered moment perpendicular to the $\rm CoO_2$ layers was
observed. Muon spin rotation ($\mu $SR) measurements confirm that
the static magnetic order encompasses nearly 100\% of the volume.
The in-plane resistivity shows a minimum around the magnetic phase
transition. The susceptibility exhibits a broad peak around 30 K,
characteristic of two-dimensional antiferromagnetic fluctuations.
These findings may be indicative of a quasi-two-dimensional
commensurate spin density wave.

Cylindrical single crystals of diameter 6 mm and length 80 mm were
grown in an optical floating-zone furnace. The initial
polycrystalline material was prepared using a mixture of $\rm
Na_2CO_3$ and $\rm Co_3O_4$ with a Na:Co ratio of 0.8:1. The
powders were calcined at 750~$^\circ$C for 12 hours and then
reacted at 850~$^\circ$C for a day with intermediate grindings.
The mixture was pressed to form a cylinder and premelted prior to
growth. The molten zone was passed through the feed rod under
oxygen flow at a rate of 2 mm/h. Details will be published
elsewhere \cite{chen}. Pieces cut from two of the resulting
crystals were analyzed by inductively coupled plasma atomic
emission spectroscopy and atomic absorption spectroscopy. The
Na:Co ratios were found to be $0.82\pm0.02$ and $0.83\pm0.02$,
respectively. The virtually identical compositions attest to good
reproducibility of the crystal preparation. The N\'{e}el
temperatures of both samples were also identical within
experimental error. Single-crystal and powder x-ray diffraction
confirm a nearly pure phase with space group P6$_3$/mmc, in good
agreement with Ref. \onlinecite{balsys96} and with Raman
scattering experiments \cite{raman}. The lattice parameters were
determined by single-crystal x-ray diffraction along $(h,h,0)$ and
$(0,0,l)$ at room temperature, resulting in values of $a=2.843(1)
{\rm \AA}$ and $c= 10.687(5) {\rm \AA}$. The susceptibility,
specific heat, and resistivity measurements were carried out in
Quantum Design MPMS and PPMS systems.

Fig. 1 shows the magnetic susceptibility measured in a 5T magnetic
field oriented either along the $c$-axis or perpendicular to it.
Upon cooling, the anisotropy increases, and a broad peak develops
around $31 \pm 3$ K. This is indicative of short-range,
Ising-type, quasi-two-dimensional antiferromagnetic correlations.
A sharp decrease of the $c$-axis susceptibility around 20 K
heralds an antiferromagnetic transition with ordered moments along
$c$. The inset in Fig. 1 shows that the transition is sharp and
only weakly affected by the field. Curie tails, presumably due to
paramagnetic impurities, are suppressed in a 5T field.

\begin{figure}
\includegraphics[width=0.75\linewidth]{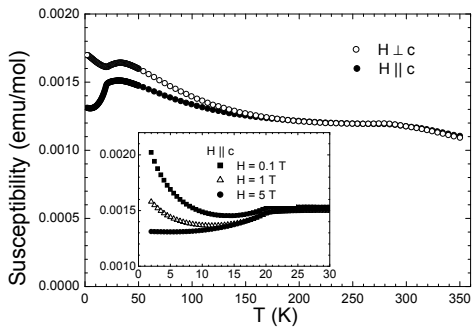}
\caption{Magnetic susceptibility of Na$_{0.82}$CoO$_2$ measured on
cooling in a 5T magnetic field oriented along (open symbols) and
perpendicular to (closed symbols) the $c$-axis. Inset:
Low-temperature susceptibility along $c$ in different applied
fields.}
\end{figure}

A broad hump in the susceptibility is evident at around $\rm 285
\pm 5 K$, for all sample orientations and at all applied magnetic
fields.  This temperature corresponds roughly to the onset of the
low-temperature splitting of a phonon mode at $\rm \nu \sim 550
cm^{-1}$, as observed in IR conductivity measurements (not shown).
This suggests that this feature in the susceptibility arises from
a structural phase transition intrinsic to $\rm Na_{0.8}CoO_2$,
rather than from an impurity phase such as CoO (an antiferromagnet
with $\rm T_{N} = 292 K$). This interpretation is consistent with
x-ray powder diffraction measurements, which exhibit no traces of
CoO. The origin of the structural transition remains to be
determined.

The specific heat, $C_P$, also exhibits a sharp anomaly at
$19.8\pm 0.1$ K, which indicates the onset of long-range
antiferromagnetic ordering (Fig. 2). The entropy contained in this
anomaly amounts to 0.08(1) J/molK and corresponds to about 10\% of
the entropy of the Co$^{4+}$ spin-1/2 system. There is a slight
excess heat capacity above $\rm T_N$, extending up to $\sim$ 35 K.
These observations are consistent with the short-range magnetic
fluctuations above $\rm T_N$ visible in the magnetic
susceptibility measurements.

\begin{figure}
\includegraphics[width=0.95\linewidth]{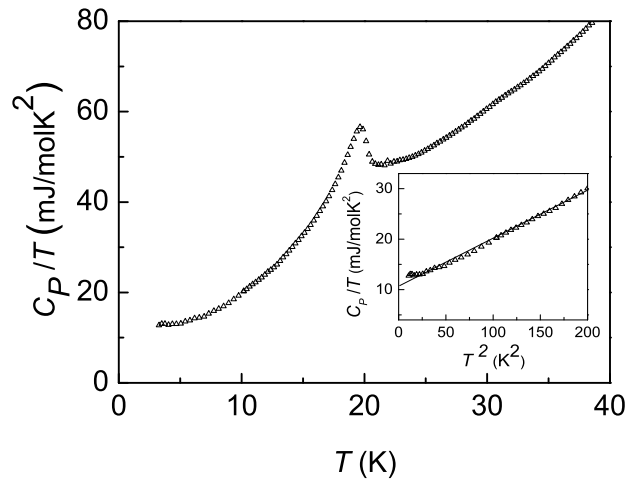}
\caption{Specific heat of a Na$_{0.82}$CoO$_2$ crystal. Inset:
Low-temperature specific heat in a
$C_P\,/\,T\,=\,\gamma\,+\,\beta\,T^2$ plot.}
\end{figure}

The specific heat at low temperatures contains a contribution
linear in temperature which is revealed in a
$C_P\,/\,T\,=\,\gamma\,+\,\beta\,T^2$ plot (inset in Fig. 2). It
is slightly sample-dependent and amounts to 8.4(3) and 10.4(4)
mJ/molK$^2$ for the two measured crystals. We ascribe it to the
Sommerfeld term from the conduction electrons. If the cubic term
$\beta$ is attributed to the contribution of acoustic phonons
alone (ignoring possible contributions from magnons), it
corresponds to a Debye temperature of $\sim$ 420 K.

Muon-spin-rotation ($\mu $SR) experiments have been performed using the GPS
setup at the $\pi $m3 beamline at the Paul-Scherrer-Institute (PSI) in
Villigen, Switzerland, which provides 100\% spin-polarized muons. The $\mu $%
SR technique is especially suited for the investigation of
magnetic materials with small magnetic moments. In particular, it
allows one to study the homogeneity of the magnetic state on a
microscopic scale, and also to access its volume fraction. The
accessible time scale is 10$^{-6}$-10$^{-10}$ s. The muons are
implanted into the bulk of the material, where they thermalize
rapidly without any noticeable loss in their initial polarization.
They stop at well-defined (but in general unknown) interstitial
lattice sites. In the present compound, they most likely form a
muoxyl bond with an oxygen atom in the CoO$_{2}$ planes or occupy
empty Na sites in between the CoO layers. The whole ensemble of
muons is randomly distributed throughout a layer with a thickness
of about 200 $\mu $m, and thus probes a representative part of the
sample volume. The muons decay (with a mean lifetime of $\tau
_{\mu }=$ 2.2 $\mu $s) into two neutrinos and a positron which is
preferentially emitted along the direction of the muon
spin at the instant of decay. The time evolution of the spin polarization, $%
P(t)$, of the muon ensemble can thus be obtained via the time-resolved
detection of the spatial asymmetry of the positron emission rate.

Fig. 3a shows $\mu $SR spectra obtained at T=7.5 K in a zero-field
configuration. The normalization of the total asymmetry of the
polarization function is based on the measurement in a weak
transverse field of 100 Oe at 50 K, where the sample is in the
paramagnetic state. Spectra for two different orientations of the
muon spin-polarization with respect to the crystallographic
$c$-axis are shown. It is evident that the spectra contain several
oscillating components, which means that the muons experience
several well-defined local magnetic fields. Even without any
further analysis, one can immediately conclude that a large
fraction of the sample volume must exhibit static magnetic order.
We analyzed the spectra using a relaxation function of the form

\begin{equation}
P(t)=P(0)\,\underset{i}{\sum }\;A_{i}\;\cos (2\pi \nu _{\mu }^{i}t)\;\exp
(-\lambda ^{i}t).
\end{equation}

An excellent description of the data can be obtained with a sum of
three oscillatory terms with frequencies and relaxation rates (at
low temperature) $\nu _{\mu }=1.25$, 2.6 and 3.2 MHz and $\lambda
=0.8,$ 1.2 and 2.8 $\mu $s$^{-1},$ plus one non-oscillatory term
with a very small relaxation rate of $\lambda =0.02$ $\mu
$s$^{-1}$ that accounts for the component of the muon spin
parallel to the internal fields (lines in Fig. 3a). The
frequencies of the oscillating components do not depend on the
direction of the muon spin, while their amplitudes, $A_{i}$,
exhibit a strong variation. The amplitudes for the different
muon-spin orientations can be used to determine the direction of
the local magnetic fields, as well as the corresponding magnetic
volume fractions. The spectra suggest that the muons experience
three different well-defined local magnetic fields which point in
different directions. The highest field is parallel to the
$c$-axis (within the error bar) and amounts to about 17\% of the
entire signal. The 2.6 MHz signal is at an angle of about
25$^{\circ }$ with respect to the $c$-axis and comprises nearly
33\% of the total amplitude. Finally, the 1.2 MHz signal is at an
angle of 43$^{\circ }$ and represents about 50\% of the entire
signal. Based on this analysis, we thus find that the volume
fraction of the magnetic state is in excess of 95\%: that is, the
entire sample is magnetically ordered within the error of our
measurement.

\begin{figure}
\includegraphics[width=0.95\linewidth]{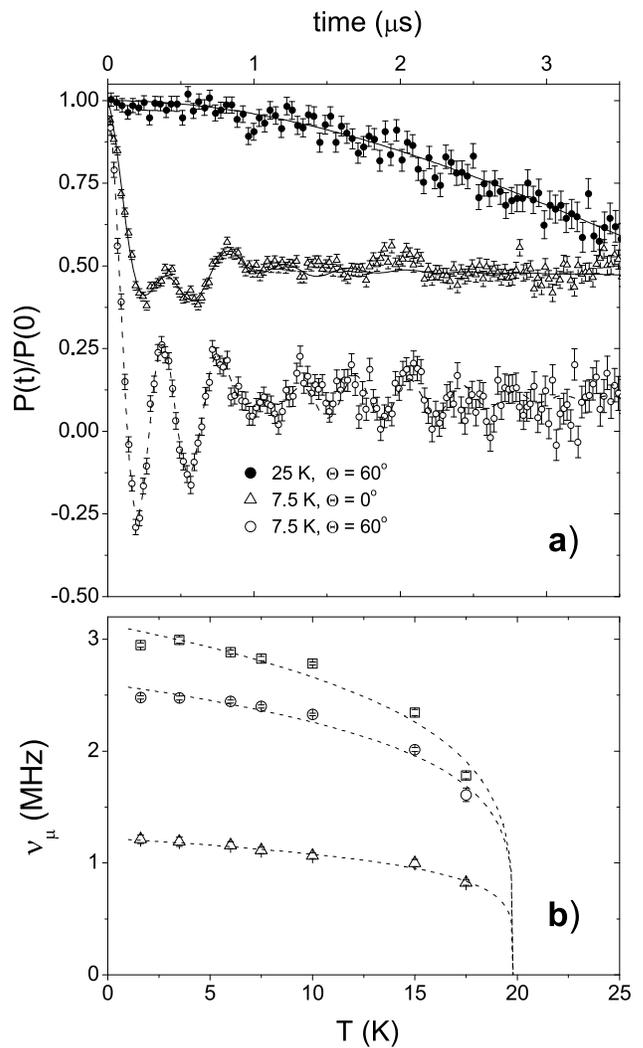}
\caption{a) Time dependence of the spin polarization of muons
implanted in a Na$_{0.82}$CoO$_2$ crystal in zero field at $\rm
T=7.5$ K. $\Theta$ is the angle subtended by the initial spin
polarization and the crystallographic $c$-axis. The lines are the
results of fits described in the text. b) Temperature dependence
of the $\mu$SR frequencies extracted from the $P(t)$ spectra
(panel a). The dashed lines are guides to the eye.}
\end{figure}

Evidence for antiferromagnetic ordering in Na$_{x}$CoO$_{2}$ was
also obtained in a previous $\mu $SR study \cite{sugiyama03}.
While the N\'{e}el temperatures we find are in good agreement with
the data of Sugiyama \textit{et al.} \cite{sugiyama03}, there are
also substantial differences. First, they obtained much smaller
volume fractions of the magnetically ordered phase (about 20\% in
their polycrystalline Na$_{0.75}$CoO$_{2}$ sample and 50\% in
their Na$_{0.9}$CoO$_{2}$ single crystal). This suggests that
their samples might have been less homogeneous, possibly due to an
inhomogeneous distribution of Na. Another important difference may
exist concerning the intercalation of water, which is known to
occur rather rapidly under moist conditions, especially for powder
samples. Secondly, the magnitude and number of the deduced muon
spin precession frequencies are also rather different. In their
polycrystalline samples, they also observe three muon frequencies,
at about 3.3, 2.6 and 2.1 MHz. However, their relative amplitudes
are different from those observed in our crystals, so that, for
instance, the largest contribution comes from the
highest-frequency mode. In their single-crystal sample, Sugiyama
\textit{et al.} observe only one frequency, though there is some
evidence for at least one additional frequency \cite{sugiyama03}.

This leaves us with the important question of whether the
different local magnetic fields are a consequence of different
crystallographic muon stopping sites, or whether they should be
interpreted in terms of different magnetic environments. The
latter may be caused by commensurate magnetic order with a large
unit cell, but could also arise from a macroscopically
inhomogeneous magnetic state. As noted by Sugiyama \textit{et
al.}, the local fields experienced by the muons in different
crystallographic muon sites should be mostly of dipolar origin.
Potential muon stopping sites are close to the oxygen ions and
near Na(1) or Na(2) vacancies. We performed a calculation of the
dipolar fields for the case of Co moments directed along the
$c$-axis and exhibiting A-type antiferromagnetic order. We find
that the magnetic fields near the Na(1) and the Na(2) vacancies
are vanishingly small. We therefore suggest that the muons are
located primarily near the oxygen ions, forming a muoxyl bond, as
in other oxide compounds including the cuprate high-$\rm T_{c}$
superconductors \cite{muonsite}. Assuming an ordered moment of 0.3
$\mu _{B}$, we obtain a local magnetic field of about 2 MHz at the
oxygen site.

The three different muon-spin precession frequencies observed in
the experiment could be due to the presence of a structural
minority phase with a monoclinic distortion, as discussed in Ref.
\onlinecite{fouassier73}. However, if a minority phase were
present in our sample, its N\'{e}el temperature would have to be
very similar to that of the majority phase. This conclusion is
based on the temperature dependence of the local magnetic fields,
which is shown in Fig 3b. It is evident that all three local
fields exhibit a similar T-dependence, i.e. they decrease towards
the magnetic transition at 19.8 K evident in the magnetization and
the specific heat data.

It is also possible that the three frequencies reflect a
commensurate spin order in which the same local field is repeated
in every third crystallographic unit cell. However, our data do
not support an interpretation in terms of an incommensurate
magnetic order \cite{sugiyama03}, which should give rise to much
larger depolarization rates and to a different form of the
relaxation function. We have tried fitting with Bessel functions,
but could not obtain a reasonable fit. Since $\mu $SR data provide
only local information, it is not straightforward to deduce
further information concerning the magnetic ordering pattern.
Neutron scattering experiments will be required to settle this
important issue.

\begin{figure}
\includegraphics[width=0.95\linewidth]{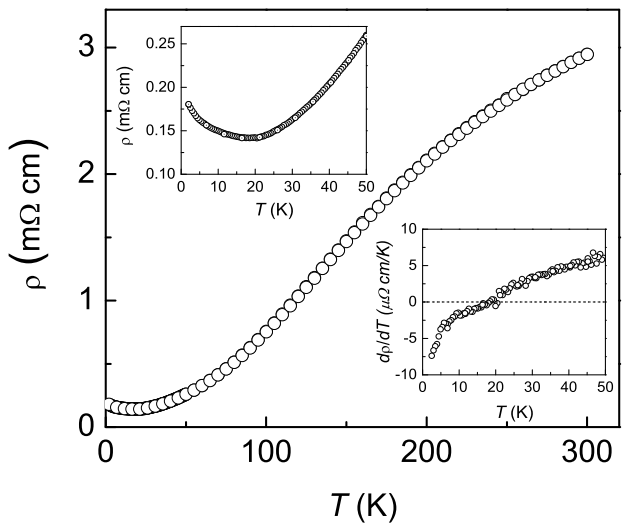}
\caption{In-plane resistivity of Na$_{0.82}$CoO$_2$. Insets:
Low-temperature resistivity and its temperature derivative.}
\end{figure}

Fig. 4 shows the in-plane resistivity measured on a rectangular
crystal (4\,$\times$\,3\,$\times$\,0.6\,mm$^3$) with a four-point
ac-technique ($\nu$\,=\,119 Hz). The resistivity is
frequency-independent between 19 and 1000 Hz and exhibits metallic
character for $\rm T>T_N$, with a residual-resistivity ratio
$\rho$\,(300K)/$\rho$\,(4.2K)\,$\sim$\,20. The temperature
derivative of the resistivity changes sign at $\rm T_N$ (insets in
Fig. 4). This is the behavior expected if a spin density wave
opens a small gap at the Fermi surface.

In conclusion, the thermodynamic, transport, and $\mu$SR data
reported here are consistent with an intrinsic, commensurate
antiferromagnetic spin density wave transition at 19.8 K, but
incompatible with incommensurate magnetic order \cite{sugiyama03}.
Our data may prove valuable in elucidating the origin of the large
thermopower of $\rm Na_{x}CoO_2$ and of the unconventional
superconducting state in the closely related compound $\rm
Na_{0.35}CoO_2 \cdot 1.3\, H_2O$.

We acknowledge the technical support of A. Amato at the PSI, and
E. Br\"ucher and G. Siegle at the MPI.

\end{document}